\documentclass[sigconf]{acmart}

\usepackage{booktabs} 
\usepackage{listings}
\usepackage{tabularx}
\usepackage{color}
\usepackage{colortbl}
\usepackage[ruled,lined,linesnumbered,vlined,algo2e]{algorithm2e} 
\usepackage{makecell}
\usepackage{changes}

\newcommand{\etc}{\mbox{\emph{etc.\ }}}

\newcommand\blfootnote[1]{%
  \begingroup
  \renewcommand\thefootnote{}\footnote{#1}%
  \addtocounter{footnote}{-1}%
  \endgroup
}

\setcopyright{none}
\renewcommand\footnotetextcopyrightpermission[1]{} 
\begin{document}
\title{Presentation Proposal: Towards Efficient Data-flow Test Data Generation Using KLEE}
\author{Chengyu Zhang\textsuperscript{1}, Ting Su\textsuperscript{2}, Yichen Yan\textsuperscript{1}, Ke Wu\textsuperscript{3}, Geguang Pu\textsuperscript{1}}
\renewcommand{\authors}{Chengyu Zhang, Ting Su, Yichen Yan, Ke Wu, Geguang Pu}
\affiliation{\textsuperscript{1}School of Computer Science and Software Engineering, East China Normal University, China} 
\affiliation{\textsuperscript{2}School of Computer Science and Engineering, Nanyang Technological University, Singapore}
\affiliation{\textsuperscript{3}National Trusted Embedded Software Engineering Technology Research Center, China}
\email{dale.chengyu.zhang@gmail.com, suting@ntu.edu.sg, sei_yichen@outlook.com}
\email{bukawu@126.com, ggpu@sei.ecnu.edu.cn}

\renewcommand{\shortauthors}{C.~Zhang, T.~Su, Y.~Yan, K.~Wu, G.~Pu}

\begin{abstract}
Dataflow coverage, one of the white-box testing criteria, focuses on the relations between variable definitions and their uses.
Several empirical studies have proved data-flow testing is more effective than control-flow testing. 
However, data-flow testing still cannot find its adoption in practice, due to the lack of effective tool support.
To this end, we propose a guided symbolic execution approach to efficiently search for program paths to satisfy data-flow coverage criteria. 
We implemented this approach on KLEE and evaluated with 30 program subjects which are constructed by the subjects used in previous data-flow testing literature, SIR, SV-COMP benchmark.

Moreover, we are planning to integrate the data-flow testing technique into the new proposed symbolic execution engine, SmartUnit, which is a cloud-based unit testing service for industrial software, supporting coverage-based testing. 
It has successfully helped several well-known corporations and institutions in China to adopt coverage-based testing in practice, totally tested more than one million lines of real code from industry.
\blfootnote{This is the presentation proposal for the 1st KLEE Workshop, 2018, London, UK.}
\end{abstract}


\maketitle
\section{Introduction}

Data-flow testing is a group of testing strategies which aims to find paths to exercise the interactions between definitions and uses of the variables.
The faults can be found by observing whether all corresponding uses produce the desired results.
The idea of data-flow testing was first proposed in 1976 by Herman~\cite{herman1976data} who claimed data-flow testing could test a program more thoroughly and reveal more subtle software bugs.
Several empirical studies have revealed that data-flow coverage criteria is more effective than control-flow coverage criteria~\cite{hutchins1994experiments, frankl1998further}.
However, data-flow coverage still cannot find its adoption in practice, due to the lack of effective tool support.

In this presentation, we propose an efficient guided testing approach which achieves data-flow coverage criteria.
To our knowledge, we are the first to adapt symbolic execution engine KLEE for data-flow testing and have implemented it on KLEE, an efficient and robust symbolic execution engine. 
To evaluate its efficiency, we build a data-flow testing benchmark, 
which consists of the subjects used in previous data-flow testing work, Software-artifact Infrastructure Repository\footnote{http://sir.unl.edu/portal/index.php} (SIR), and International Competition on Software Verification (SV-COMP) benchmark. 

\section{Approach and Implementation}

\subsection{Approach}
According to the Herman's definition~\cite{herman1976data}, a \emph{def-use pair} exists when there is at least one program \emph{path} from the \emph{definition} of variable to the \emph{use}
where there are no redefinitions between \emph{definition} and \emph{use}. For a \emph{def-use pair}, if there is an input \emph{t} that induces an execution path passing through \emph{definition} and then \emph{use} with no intermediate 
redefinitions of \emph{x} between \emph{definition} and \emph{use}, the input \emph{t} satisfies the \emph{def-use pair}. The requirement to cover all \emph{def-use pairs} at least once is called \emph{def-use coverage criterion}.

\begin{figure}
    \includegraphics[width=0.4\textwidth]{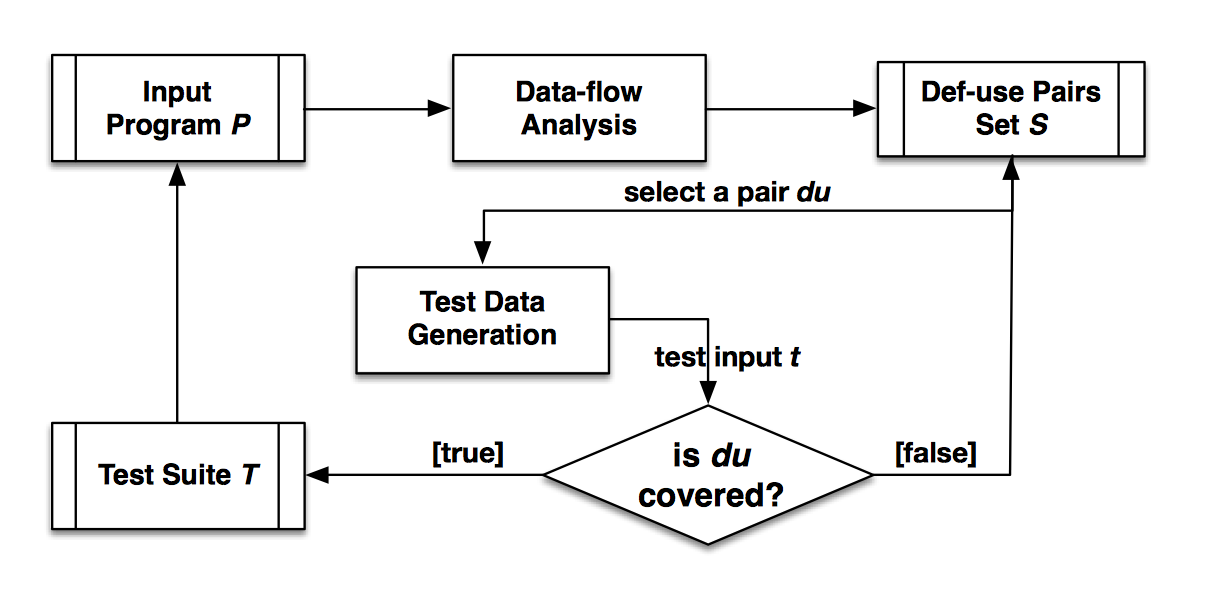}
    \caption{The basic process of data-flow testing.}
    \label{fig:process}
\end{figure}

Figure.~\ref{fig:process} shows the basic process of data-flow testing. 
For an input program, we perform static analysis to get a set of def-use pairs.
At each time, the test generator selects a pair as the target to find a test input covering the pair.
If the generator successfully generates the test input covering the pair, the test input will be put in the test suite, otherwise, the pair is regarded as the unsatisfiable pair within the given testing time.

To achieve efficient data-flow testing, we designed a \emph{cut-points guided search algorithm} to enhance symbolic execution~\cite{su2015combining}. The key idea of the strategy is to reduce unnecessary path exploration and provide more guidance during execution.
There are three key elements used to optimize the strategy: \emph{Cut Points}, \emph{Instruction Distance} and \emph{Redefinition Pruning}.
(1) \emph{Cut Points} is a sequence of control points that must be traversed through by any path that could cover the target pair. 
These cut points are used as intermediate goals during the search to narrow down the exploration space of symbolic execution.
(2) \emph{Instruction Distance} is the distance between currently executed instruction and target instruction in the control flow graph. 
The execution state which has shorter instruction distance toward the goal can reach it more easily.
(3) \emph{Redefinition Pruning} is the strategy that gives lower priority to the execution states which have the redefinitions on paths.
According to the definition of \emph{covering def-use pair}, the path which has redefinitions between \emph{definition} and \emph{use} is invalid.
Pruning the potential invalid paths can avoid useless explorations.

\subsection{Implementation}
Our implementation follows the basic data-flow testing process. 
In the data-flow analysis phase,
the input program is analyzed by CIL~\cite{necula2002cil} tool which is an infrastructure for C program analysis and transformation to identify the def-use pairs, cut points, and static program information. 
We use CIL instead of LLVM because we intend to analyze def-use pairs from source-level instead of bytecode level (LLVM IR).
The test data generation and coverage computation phase are implemented on KLEE. 

In detail, we implemented a searcher class in KLEE to apply our cut-points guided search strategy. 
Furthermore, we constructed a data-flow information table in KLEE to supply some extra data-flow information such as redefinitions and cut points during execution. 
A coverage checker and a redefinition checker are also implemented in KLEE based on the process tree. 
If an execution state reaches the use through the definition without any redefinition on the path, 
the coverage checker identifies that it has covered the pair.
Then the execution state generates the test case satisfying the pair.
\section{evaluation and discussion} 

Although data-flow testing has been investigated for a long time, there are still no standard benchmark programs for evaluating data-flow testing techniques.
To set up a fair evaluation basis, we constructed our benchmarks as follows: (1) We conducted a survey on previous data-flow testing literature~\cite{su2017survey}. 
After excluding the subjects that are not available, not written in C language and simple laboratory programs, 
we finally got 7 subjects. (2) The benchmark included 7 Simens subjects from SIR which are widely used in the experiments of program analysis and software testing. 
(3) We further enriched the repository with 16 distinct subjects from SV-COMP which is used in the competition for software verification. We chose two groups of subjects in SV-COMP benchmark, \textsf{ntdriver group} (with 6 subjects) and \textsf{ssh} group (with 10 subjects).
The code and pair scale of each categories are showed in Table.~\ref{table:tools}.

\begin{table}[!t]
\scriptsize
\newcommand{\tabincell}[2]{\begin{tabular}{@{}#1@{}}#2\end{tabular}}
\renewcommand{\arraystretch}{1}
\caption{Evaluation Statistic of Data-flow Testing via KLEE}
\label{table:tools}
\centering
\begin{tabular}{|c||c|c|c|c|c|c|c|}
\hline
Subjects &\#Sub &\#LOC &\#DU pair &\multicolumn{2}{c|}{Average Coverage} &\multicolumn{2}{c|}{\makecell{Median Time \\ (s/pair)}}\\
\hline
& & & &KLEE &CPA &KLEE &CPA \\
Previous Literature &7 &449 &346  &60\% &72\% &0.1 &4.3\\ 
SIR &7 &2,687 &1,409 &57\% &60\% &0.7 &10.1\\
SV-COMP \textsf{(ntdriver)} &6 &7,266 &2,691 &75\% &51\% &1.5 &5\\
SV-COMP \textsf{(ssh)} &10 &5,249 &18,347 &29\% &31\% &18 &5.7\\
\hline
\end{tabular}
\end{table}
Table.~\ref{table:tools} gives an overview of our evaluation. It shows the data-flow coverage (The ratio between the number of covered pairs and the total number of pairs) and median test time of KLEE and CPAchecker approach~\cite{su2015combining}.
From Table~\ref{table:tools}, we can find that KLEE can easily achieve nearly 60\% of data-flow coverage within less than 1 second for each pair in the subjects from previous literature and SIR subjects.
Comparing with the model-checking approach implemented on CPAchecker, it spends less time and achieves the similar coverage in most of the subjects.
We found the longer median time in SV-COMP \textsf{ssh} benchmark is caused by the its complexity, which have complicate loops.
The existence of a large number of infeasible pairs is also a obstacle for applying symbolic execution in practice, since it will waste much time on useless exploration.
To solve this problem, we use model-checking to filter out the infeasible pairs in order to keep symbolic execution away from useless explorations.
\section{Application}

The idea in this presentation comes from our previous work published in ICSE'15~\cite{su2015combining}. According to our survey on data-flow testing~\cite{su2017survey}, there are a variety of approaches to generate data-flow test data, such as \emph{random testing}, \emph{collateral coverage-based testing}, \emph{search-based testing} and
\emph{model checking-based testing}. However, to our knowledge, we are the first to adapt symbolic execution for data-flow testing. Symbolic execution is a more efficient and precise way, because it can straightforwardly find the paths that satisfy the data-flow coverage criteria and easily generate the test case.

Furthermore, we implemented a cloud-based industrial automated unit test generation framework named SmartUnit~\cite{zhang2018smartunit} which
 depends on our previous work on symbolic execution~\cite{su2014automated,su2016automated}.
It helps several corporations and institutions to adopt coverage-based testing in practice, include \emph{China Academy of Space Technology}, the main spacecraft development and production agency in China (like NASA in the United States);
\emph{CASCO Signal Ltd.}, the best railway signal technique corporation in China, \etc.
SmartUnit has totally tested more than one million lines of code since its release in 2016.
\section{conclusion}

In this presentation, we propose an efficient data-flow test data generation algorithm implemented on KLEE and evaluated on a diverse set of program subjects.
It enables efficient and effective data-flow testing and helps several corporations and institutions to adopt data-flow testing in practice.

\bibliographystyle{ACM-Reference-Format}
\bibliography{sample-bibliography} 


\begin{thebibliography}{9}


\ifx \showCODEN    \undefined \def \showCODEN     #1{\unskip}     \fi
\ifx \showDOI      \undefined \def \showDOI       #1{#1}\fi
\ifx \showISBNx    \undefined \def \showISBNx     #1{\unskip}     \fi
\ifx \showISBNxiii \undefined \def \showISBNxiii  #1{\unskip}     \fi
\ifx \showISSN     \undefined \def \showISSN      #1{\unskip}     \fi
\ifx \showLCCN     \undefined \def \showLCCN      #1{\unskip}     \fi
\ifx \shownote     \undefined \def \shownote      #1{#1}          \fi
\ifx \showarticletitle \undefined \def \showarticletitle #1{#1}   \fi
\ifx \showURL      \undefined \def \showURL       {\relax}        \fi
\providecommand\bibfield[2]{#2}
\providecommand\bibinfo[2]{#2}
\providecommand\natexlab[1]{#1}
\providecommand\showeprint[2][]{arXiv:#2}

\bibitem[\protect\citeauthoryear{Frankl and Iakounenko}{Frankl and
  Iakounenko}{1998}]%
        {frankl1998further}
\bibfield{author}{\bibinfo{person}{Phyllis~G Frankl} {and}
  \bibinfo{person}{Oleg Iakounenko}.} \bibinfo{year}{1998}\natexlab{}.
\newblock \showarticletitle{Further empirical studies of test effectiveness}.
\newblock \bibinfo{journal}{{\em ACM SIGSOFT Software Engineering Notes\/}}
  \bibinfo{volume}{23}, \bibinfo{number}{6} (\bibinfo{year}{1998}),
  \bibinfo{pages}{153--162}.
\newblock


\bibitem[\protect\citeauthoryear{Herman}{Herman}{1976}]%
        {herman1976data}
\bibfield{author}{\bibinfo{person}{PM Herman}.}
  \bibinfo{year}{1976}\natexlab{}.
\newblock \showarticletitle{A data flow analysis approach to program testing}.
\newblock \bibinfo{journal}{{\em Australian Computer Journal\/}}
  \bibinfo{volume}{8}, \bibinfo{number}{3} (\bibinfo{year}{1976}),
  \bibinfo{pages}{92--96}.
\newblock


\bibitem[\protect\citeauthoryear{Hutchins, Foster, Goradia, and
  Ostrand}{Hutchins et~al\mbox{.}}{1994}]%
        {hutchins1994experiments}
\bibfield{author}{\bibinfo{person}{Monica Hutchins}, \bibinfo{person}{Herb
  Foster}, \bibinfo{person}{Tarak Goradia}, {and} \bibinfo{person}{Thomas
  Ostrand}.} \bibinfo{year}{1994}\natexlab{}.
\newblock \showarticletitle{Experiments on the effectiveness of dataflow-and
  control-flow-based test adequacy criteria}. In \bibinfo{booktitle}{{\em
  ICSE}}. IEEE, \bibinfo{pages}{191--200}.
\newblock


\bibitem[\protect\citeauthoryear{Necula, McPeak, Rahul, and Weimer}{Necula
  et~al\mbox{.}}{2002}]%
        {necula2002cil}
\bibfield{author}{\bibinfo{person}{George Necula}, \bibinfo{person}{Scott
  McPeak}, \bibinfo{person}{Shree Rahul}, {and} \bibinfo{person}{Westley
  Weimer}.} \bibinfo{year}{2002}\natexlab{}.
\newblock \showarticletitle{CIL: Intermediate language and tools for analysis
  and transformation of C programs}. In \bibinfo{booktitle}{{\em Compiler
  Construction}}. Springer, \bibinfo{pages}{209--265}.
\newblock


\bibitem[\protect\citeauthoryear{Su, Fu, Pu, He, and Su}{Su
  et~al\mbox{.}}{2015}]%
        {su2015combining}
\bibfield{author}{\bibinfo{person}{Ting Su}, \bibinfo{person}{Zhoulai Fu},
  \bibinfo{person}{Geguang Pu}, \bibinfo{person}{Jifeng He}, {and}
  \bibinfo{person}{Zhendong Su}.} \bibinfo{year}{2015}\natexlab{}.
\newblock \showarticletitle{Combining symbolic execution and model checking for
  data flow testing}. In \bibinfo{booktitle}{{\em ICSE}}.
  \bibinfo{pages}{654--665}.
\newblock


\bibitem[\protect\citeauthoryear{Su, Pu, Fang, He, Yan, Jiang, and Zhao}{Su
  et~al\mbox{.}}{2014}]%
        {su2014automated}
\bibfield{author}{\bibinfo{person}{Ting Su}, \bibinfo{person}{Geguang Pu},
  \bibinfo{person}{Bin Fang}, \bibinfo{person}{Jifeng He}, \bibinfo{person}{Jun
  Yan}, \bibinfo{person}{Siyuan Jiang}, {and} \bibinfo{person}{Jianjun Zhao}.}
  \bibinfo{year}{2014}\natexlab{}.
\newblock \showarticletitle{Automated coverage-driven test data generation
  using dynamic symbolic execution}. In \bibinfo{booktitle}{{\em SERE}}. IEEE,
  \bibinfo{pages}{98--107}.
\newblock


\bibitem[\protect\citeauthoryear{Su, Pu, Miao, He, and Su}{Su
  et~al\mbox{.}}{2016}]%
        {su2016automated}
\bibfield{author}{\bibinfo{person}{Ting Su}, \bibinfo{person}{Geguang Pu},
  \bibinfo{person}{Weikai Miao}, \bibinfo{person}{Jifeng He}, {and}
  \bibinfo{person}{Zhendong Su}.} \bibinfo{year}{2016}\natexlab{}.
\newblock \showarticletitle{Automated coverage-driven testing: combining
  symbolic execution and model checking}.
\newblock \bibinfo{journal}{{\em {SCIENCE} {CHINA} Information Sciences\/}}
  \bibinfo{volume}{59}, \bibinfo{number}{9} (\bibinfo{year}{2016}),
  \bibinfo{pages}{98101}.
\newblock


\bibitem[\protect\citeauthoryear{Su, Wu, Miao, Pu, He, Chen, and Su}{Su
  et~al\mbox{.}}{2017}]%
        {su2017survey}
\bibfield{author}{\bibinfo{person}{Ting Su}, \bibinfo{person}{Ke Wu},
  \bibinfo{person}{Weikai Miao}, \bibinfo{person}{Geguang Pu},
  \bibinfo{person}{Jifeng He}, \bibinfo{person}{Yuting Chen}, {and}
  \bibinfo{person}{Zhendong Su}.} \bibinfo{year}{2017}\natexlab{}.
\newblock \showarticletitle{A Survey on Data-Flow Testing}.
\newblock \bibinfo{journal}{{\em CSUR\/}} \bibinfo{volume}{50},
  \bibinfo{number}{1} (\bibinfo{year}{2017}), \bibinfo{pages}{5}.
\newblock


\bibitem[\protect\citeauthoryear{Zhang, Yan, Zhou, Yao, Wu, Su, Miao, and
  Pu}{Zhang et~al\mbox{.}}{2018}]%
        {zhang2018smartunit}
\bibfield{author}{\bibinfo{person}{Chengyu Zhang}, \bibinfo{person}{Yichen
  Yan}, \bibinfo{person}{Hanru Zhou}, \bibinfo{person}{Yinbo Yao},
  \bibinfo{person}{Ke Wu}, \bibinfo{person}{Ting Su}, \bibinfo{person}{Weikai
  Miao}, {and} \bibinfo{person}{Geguang Pu}.} \bibinfo{year}{2018}\natexlab{}.
\newblock \showarticletitle{SmartUnit: Empirical Evaluations for Automated Unit
  Testing of Embedded Software in Industry}. In \bibinfo{booktitle}{{\em
  ICSE-SEIP}}. \bibinfo{pages}{10 pages}.
\newblock


\end{thebibliography}

\end{document}